# Thermal conductance across bonded SiO$_x$-SiO$_x$ interfaces in hybrid bonding process

Xingqiang Zhang[1,+], Liu Chang[2,+], Liyi Li[2,*], Zhe Cheng[1,*]

[1] School of Integrated Circuits and Beijing Advanced Innovation Center for Integrated Circuits, Peking University, Beijing 100871, China

[2] School of Integrated Circuits, Southeast University, Nanjing, 210000, China

[+] These authors contributed equally

[*] Authors to whom correspondence should be addressed: zhe.cheng@pku.edu.cn; liyi_li@seu.edu.cn

**Abstract**

Hybrid bonding is a pivotal technology for enabling three-dimensional integrated circuits (3D-ICs). Among the foremost challenges facing 3D-IC implementation is thermal management, where a deep understanding of heat conduction across bonded interfaces is essential for addressing heat dissipation and reliability issues. Nevertheless, the thermal conductance of bonded dielectric–dielectric interfaces remains poorly understood. In this study, we employ the low-temperature bonding technique integral to hybrid bonding to fabricate SiO$_x$–SiO$_x$ interfaces and investigate their thermal boundary conductance (TBC) using time-domain thermoreflectance (TDTR). Structural characterizations show high-quality bonded interfaces. By fitting the data with an equivalent multilayer thermal model, we establish a lower-limit TBC of 150 MW m$^{-2}$ K$^{-1}$ for the SiO$_x$–SiO$_x$ interfaces, which corresponds to a thermal resistance lower than that of a 9.2-nm-thick dielectric layer. These findings offer valuable insights into thermal transport in hybrid-bonded structures and provide critical guidance for the thermal design of advanced packaging solutions.

Three-dimensional integrated circuits (3D-ICs) is promising to extend Moore's Law.[1-3] Advanced packaging technologies such as hybrid bonding are critical to achieving high-density interconnects for the heterogeneous integration and interconnection of chips with different manufacturing processes.[4] By high-precision alignment of Cu vias and surrounding dielectric materials (e.g., SiO$_x$), hybrid bonding achieves ultra high density electrical interconnections comparable to monolithic integration performance.[5-8] However, to meet the stringent thermal budget requirements of advanced packaging processes, low-temperature hybrid bonding has become an indispensable technology.[9, 10] Although plasma activation facilitates the formation of interfacial covalent bonds to enable bonding at low temperatures, it may also lead to suboptimal interfacial contact, such as van der Waals-like weak interface, at the dielectric-dielectric bonding interface.[11] Such inferior contact can introduce excessively high thermal resistance, which poses a critical challenge to the heat dissipation of the entire packaging system.

As critical signal routes for vertically stacked dies, bonded interconnects which provide high I/O density can also act as heat sources.[12] Meanwhile, as a key component of the heat dissipation path, they generate hotspots, thermal gradients, and thermal strains at specific locations during chip operation, leading to performance degradation of metal pathways and bonding regions.[13-16] Thus, investigating the thermal resistance distribution in these regions is crucial, and the thermal properties of Cu vias and the surrounding dielectrics warrant accurate characterization and optimization. The processing technology and materials differ significantly from those of conventional high-temperature bonding technologies, especially in hybrid bonding processes designed to meet the low thermal budget requirements of chip manufacturing.

Kim and Hwang et al investigated the thermal transport properties of Cu-Cu bonded interfaces with passivation metals.[17] The thermal properties of Cu-dielectric interfaces have also been studied.[18-20] However, the thermal conductance of the dielectric-dielectric interfaces is much less studied due to the difficulties in thermal measurements, even though these interfaces may exhibit low thermal boundary conductance (TBC) due to voids and van der Waals contacts at the interfaces. The large thermal resistance in hybrid bonded interfaces would degrade the performance of 3D-ICs. Therefore, investigating the thermal properties of the dielectric-dielectric interfaces is not only critical for quantifying thermal resistance distribution at packaging level but also provides key guidance for optimizing low-temperature bonding processes, specifically, how to avoid weak physical contact and eliminate interfacial voids to achieve robust and strong bonding.

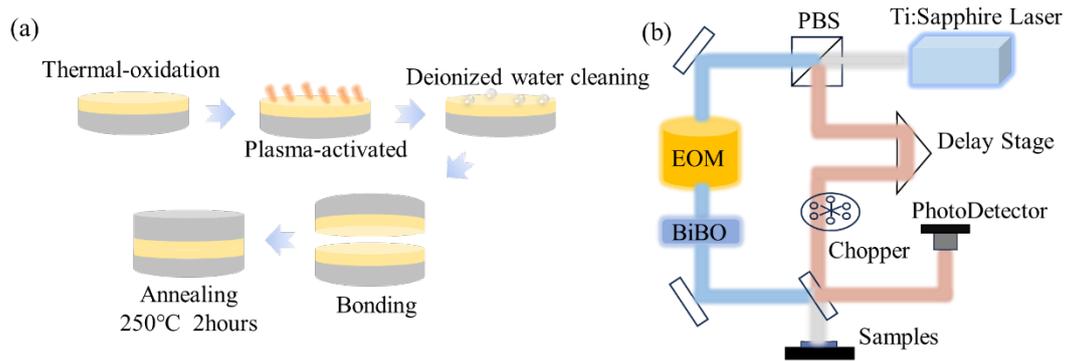

Figure 1. (a) Schematic diagram of fabrication of $SiO_x$-$SiO_x$ wafer bonded structure (b) Schematic diagram of the TDTR measurement.

In this work, we fabricated a $SiO_x$-$SiO_x$ interface via plasma activation and low-temperature (250℃) bonding. The $SiO_x$ thin films (≈ 500 nm) were grown on Si substrates through thermal oxidation in an oxygen environment at 800°C, ensuring high film quality and uniformity. Figure 1(a) illustrates the detailed fabrication process. Prior to bonding, plasma activation was performed on the sample surfaces. After plasma activation, the samples were rinsed with deionized (DI) water for 10 s and subsequently dried with nitrogen gas. Following the bonding process, thermal annealing was conducted at 250 °C for 2 h.

We employed time-domain thermoreflectance (TDTR) technology to characterize the thermal properties of the bonded samples. As shown in Fig. 1(b), a femtosecond pulsed laser (800nm) is split into a pump beam and a probe beam via a polarized beam splitter (PBS). The pump laser is amplitude-modulated by an electro-optical modulator (EOM) and then frequency-doubled (400nm) using a bismuth borate ($BiB_3O_6$, BIBO) crystal, serving as a heat source on the sample surface. In contrast, the probe beam achieves different delay times relative to the pump laser through a mechanical delay stage, which is used to extract the temperature decay after thermal excitation. The thermal penetration depth of TDTR ranges from tens of nanometers to micrometers depending on the thermal diffusivity of the materials and heating frequency: $d_p = \sqrt{\alpha / \pi f_{mod}}$, where $\alpha$ is thermal diffusivity and $f_{mod}$ is modulation frequency of pump beam. To characterize the $SiO_x$- $SiO_x$ interface of interest, the $SiO_x$ layer on one side of the interface needs to be thinner than the thermal penetration depth. This ensures that the thermal gradient can reach the interface. Based on the thermal properties of amorphous $SiO_x$ and the modulation frequencies employed in this work, the thermal penetration depth is estimated to range from 165 nm (at 10 MHz) to 280 nm (at 3.37 MHz).

For the bonded structure, the $SiO_x$ layers have a thickness of approximately 500 nm which is thermally thick for TDTR measurements of the interface. Thus, we employed chemical mechanical polishing (CMP) to thin the upper $SiO_x$ layer, as illustrated in Fig. 2. The sample was tilted at a very small angle during CMP to expose all the layers of interest and the $SiO_x$-$SiO_x$ interface. The samples were cleaned with anhydrous ethanol and DI water to eliminate organic residues generated by CMP on the exposed cross-sections. Subsequently, a thin aluminum film ($\approx$ 80 nm) was deposited on the sample surface as transducer. By focusing the laser spot at different positions and observing the echoes in the in-phase signal curve, the thickness of the upper $SiO_x$ layer was determined by the picosecond acoustic technique.[21] Herein, we selected different thicknesses of the upper $SiO_x$ layer (all around 100 nm) at various modulation frequencies to ensure our measurements penetrate through the bonded interfaces.

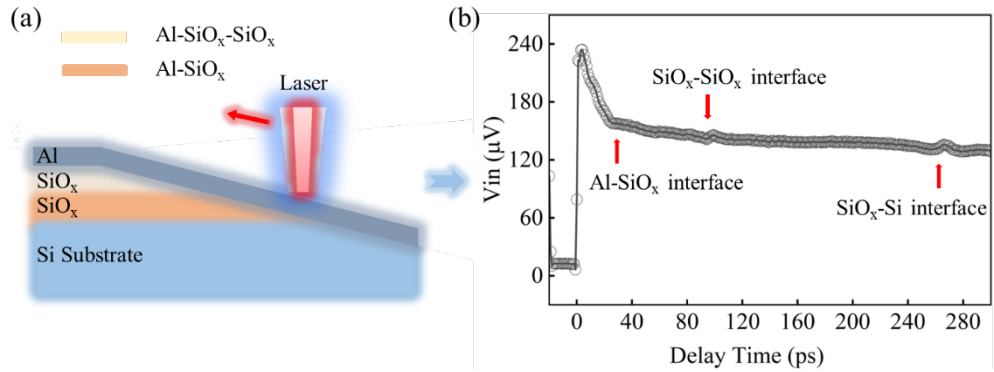

Figure 2. The schematic diagram of the polished sample (a) and the echoes corresponding to different interfaces (b).

The thermal properties measured by TDTR (including thermal conductivity and TBC ) were obtained by fitting the experimentally measured TDTR ratio signal to the analytical solutions of a multilayer thermal conduction model.[22] The thickness of each layer in the multilayer structure needs to be determined in advance. The bottom 500-nm-thick $SiO_x$ layer was thermally thick and thus treated as a bulk substrate in the data fitting. We selected the middle point of the rising peak of the $SiO_x$-$SiO_x$ signal as the interface position. The thickness of the upper $SiO_x$ layer is the product of the difference of echo positions of the $SiO_x$-$SiO_x$ interface and the Al-$SiO_x$ interface and the longitudinal sound velocity of $SiO_x$ ($\approx$ 5800 ms$^{-1}$), then divided by two.

Two sets of pump/probe laser spot radii were employed in our experiments (15 μm/7.5

μm and 6.3 μm /6.3 μm). At each tested modulation frequency, an appropriate position on the sample was selected according to the echo positions to ensure that the thickness of the upper $SiO_x$ layer was approximately 100 nm and the TDTR heating can penetrate through the bonded interfaces. The ratio curves were then recorded accordingly, with detailed parameters listed in Table 1.

Table 1. Parameters of TDTR measurements.

| Pump/probe radii (μm) | Frequency (MHz) | Thermal penetration depth (nm) | Thickness of upper $SiO_x$ (nm) |
|---|---|---|---|
| 15/7.5 | 10 | 165 | 99 |
| 15/7.5 | 5.63 | 219 | 96 |
| 6.3/6.3 | 5.02 | 232 | 84 |
| 15/7.5 | 3.37 | 283 | 136 |

For the sample structure adopted in this work, the multilayered configuration involves multiple thermal properties, including the TBC of the $Al$-$SiO_x$ interface, the TBC of the $SiO_x$-$SiO_x$ interface (the interface of interest), and the thermal conductivities of the upper and bottom $SiO_x$ layers. Prior to the measurements at the positions listed in Table 1, TDTR measurements (10 MHz) were performed on regions of the sample that do not contain the $SiO_x$-$SiO_x$ interface. These measurements were used to measure the TBC of the $Al$-$SiO_x$ interface and the thermal conductivity of $SiO_x$, with both values fitted simultaneously. The fitted TBC of the Al-SiO$_x$ interface was 117 MW m$^{-2}$ K$^{-1}$, and the thermal conductivity of SiO$_x$ was 1.36 W m$^{-1}$ K$^{-1}$ (with an 8% error) which is consistent with the reported values in the literature.[23] Since the upper and bottom SiO$_x$ layers were fabricated via the same process, and the thickness of the upper SiO$_x$ layer at all selected positions in Table 1 is approximately 100 nm (a thickness free of the size effect for SiO$_x$), we reasonably assumed that the thermal conductivities of the upper and bottom SiO$_x$ layers are equal and fixed at this fitted value.

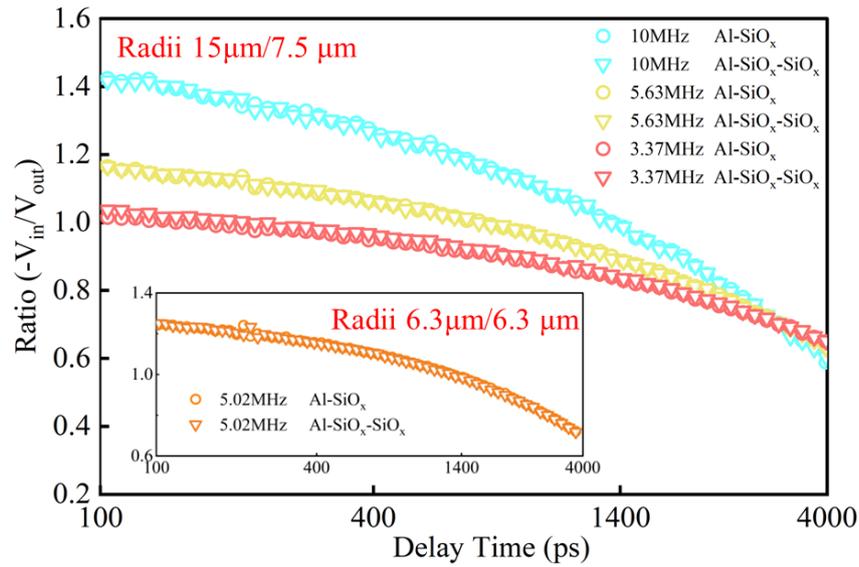

Figure 3. TDTR ratio data of Al-SiO$_x$ sample structure and Al-SiO$_x$-SiO$_x$ sample structure with different modulation frequencies and pump/probe radii.

Figure. 3 shows the results of the TDTR measurements. At the same modulation frequency, two separate measurements were performed: one was conducted on sample regions with the SiO$_x$-SiO$_x$ interface included in the structure, and the other on regions with only the Al-SiO$_x$ bilayer structure. These comparative measurements were implemented to investigate the effect of the $SiO_x$-$SiO_x$ interface on the ratio curves, thereby evaluating the effect of the introduced interfacial thermal resistance (ITR, the reciprocal of TBC) at the $SiO_x$-$SiO_x$ interface. The results demonstrate that there is almost no discernible difference in the measured signals with and without the $SiO_x$-$SiO_x$ interfacial thermal resistance. This sufficiently indicates that the interface formed via plasma activation and subsequent low-temperature bonding annealing exhibits a considerably low interfacial thermal resistance. It is challenging to extract the TBC of the interface between two low thermal conductivity layers, because the additional interfacial thermal resistance introduced by the $SiO_x$-$SiO_x$ interface corresponds to an extremely short Kapitza length. Specifically, based on the thermal conductivity of $SiO_x$ (1.38 W m$^{-1}$ K$^{-1}$), a TBC of 150 MW m$^{-2}$ K$^{-1}$ for this interface corresponds to an equivalent $SiO_x$ thickness of only 9.2 nm. After low-temperature annealing, Si–O–Si covalent bonds form at the $SiO_x$ interface, which significantly improves the properties of this homogenous interface and thus yields a considerably high TBC (the Kapitza length for the $SiO_x$-$SiO_x$ interface in this work is less than 10 nm), which has a negligible influence on the overall thermal resistance.

To gain further insight into the interfacial properties, high-resolution transmission electron microscopy (HRTEM) and energy-dispersive X-ray spectroscopy (EDS) characterizations were performed on the bonded interfaces.

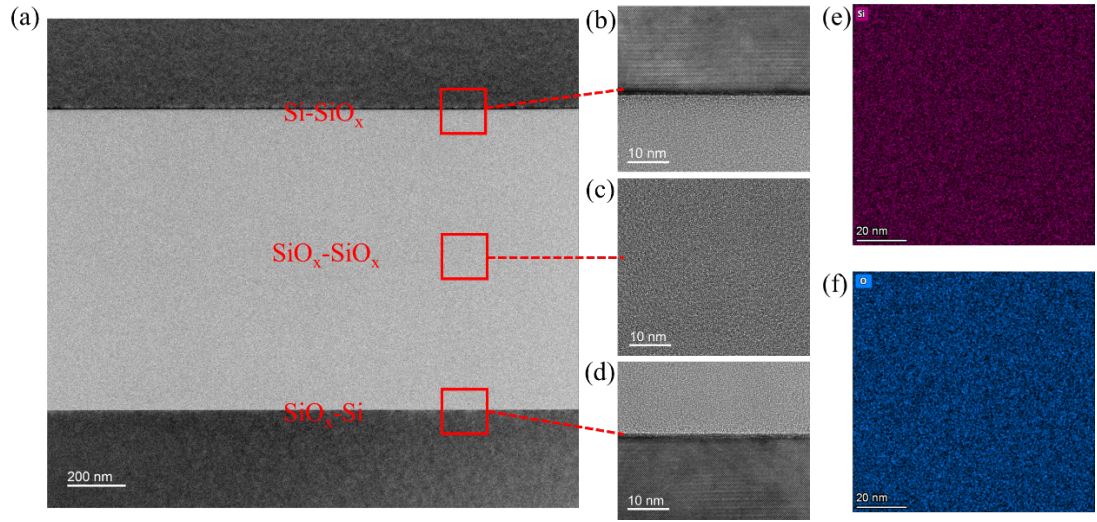

Figure 4. Transmission Electron Microscope (TEM) and Energy Dispersive X-ray Spectroscopy (EDS) images. (a) Overall structure image of sample; (b)~(d) Si-SiO$_x$ and SiO$_x$-SiO$_x$ interfaces; (e)~(f) silicon (Si) and oxygen (O) elemental distributions.

As shown in Fig. 4(a), the bonded structure comprises both Si-SiO$_x$ and SiO$_x$-SiO$_x$ interfaces. The SiO$_x$ dielectric layers were prepared via thermal oxidation, a process known to produce high-density amorphous films with good uniformity compared to deposited oxides. For the Si-SiO$_x$ interface, the two constituent materials exhibit distinct crystal structures, which results in a very sharp and well-defined interface in the TEM images (Figs. 4(b) and (d)). In contrast, the SiO$_x$-SiO$_x$ bonded interface, formed between two identical amorphous layers, is indistinguishable. Even under HRTEM imaging (Fig. 4(c)), no structural discontinuity can be identified, as shown in Fig. 4(c). After plasma activation, bonding, and low-temperature annealing, covalent bonds (Si–O–Si bonds) are formed at the interfaces.[24] Figures 4(e) and (f) show the distributions of the two key elements (Si and O) across the interfacial region. The distribution is uniform, with no discernible differences between the two sides of the interface. This indicates that there is no damage, no severe structural defects (such as voids ) at the interface after bonding and annealing. For the sample structure employed in this work, we therefore conclude that the $SiO_x$ layer with relatively low thermal conductivity forms a nearly perfect bonding interface. The $SiO_x$-$SiO_x$ interface can be regarded as being buried within the bulk $SiO_x$ layer, such that its corresponding equivalent Kapitza length is too thin to be measured exactly via TDTR measurements. For hybrid bonding structures, the interfacial thermal resistance introduced by the dielectric-dielectric interface can thus be neglected with high-quality bonding process.

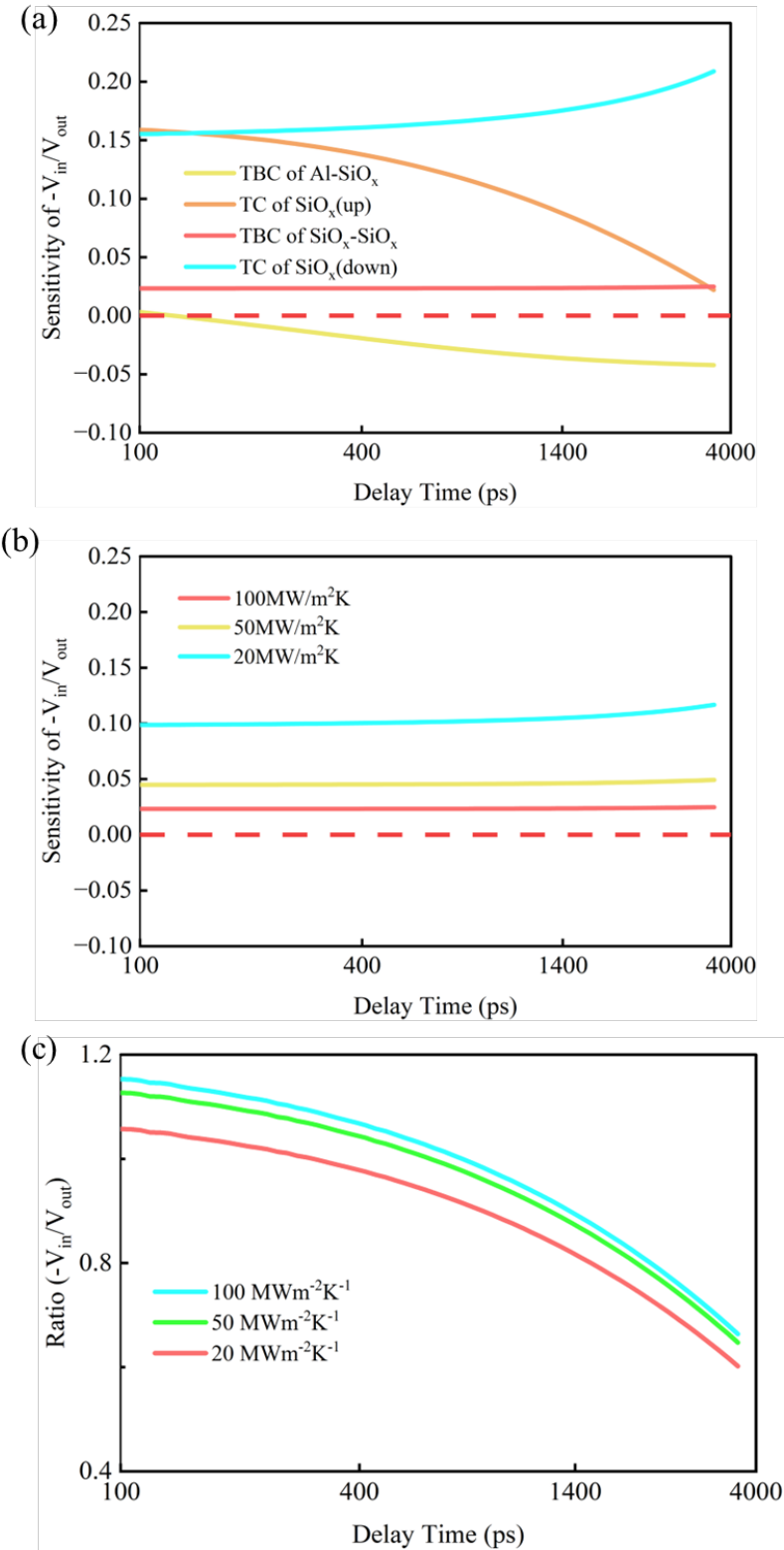

Figure 5. (a) TDTR sensitivity analysis with the pump/probe radii of 15 and 7.5 μm respectively, and modulation frequency of 5.63MHz. The thermal conductivities of Al and SiO$_x$ are fixed at 170 W m$^{-1}$ K$^{-1}$ and 1.38 W m$^{-1}$ K$^{-1}$, respectively, and the TBC

values of Al-SiO$_x$ and SiO$_x$-SiO$_x$ interfaces are both set to 100 MW m$^{-2}$ K$^{-1}$; (b) Sensitivity of the SiO$_x$-SiO$_x$ TBC with values of 100, 50, and 20 MW m$^{-2}$ K$^{-1}$ while keeping other parameters the same. (c) Calculated ratio data under the same parameter settings as in (b).

Figure 5(a) presents the calculated sensitivity of different key parameters at the given initial values, where the sensitivity of the SiO$_x$-SiO$_x$ interface at a TBC of 100 MW m$^{-2}$ K$^{-1}$ is lower than that of all other parameters. As shown in Fig. 5(b), a lower SiO$_x$-SiO$_x$ TBC corresponds to a higher sensitivity. This is because a lower TBC indicates a larger interfacial thermal resistance introduced at the interface, which imposes a more significant influence and thus results in a higher sensitivity. The ratio data (Fig. 5(c)) exhibit distinct differences for the different initial values of SiO$_x$-SiO$_x$ TBC which indicates that the TBC of interest in our experiments is sufficiently high, which is unable to induce measurable differences in the acquired ratio curves. In other words, the ratio curves show almost no discrepancy whether the SiO$_x$-SiO$_x$ interface is included in the structure or not. The ratio curves in Fig. 3 reflect that the SiO$_x$-SiO$_x$ interface of interest exhibits a considerably high TBC.

To obtain an equivalent TBC value for the SiO$_x$-SiO$_x$ interface, the thermal conductivity of SiO$_x$ was fixed at a series of values ranging from 1.25 to 1.5 W m$^{-1}$ K$^{-1}$, while the TBC values of both the Al-SiO$_x$ and SiO$_x$-SiO$_x$ interfaces were fitted simultaneously. The corresponding results are presented in Figs. 6(a)–(d). The thermal conductivity of

SiO$_x$ can be considered to lie in the range of 1.25 to 1.38 W m$^{-1}$ K$^{-1}$. The equivalent fitting results in Fig. 6 show that the TBC values of the Al-SiO$_x$ interface are all around 100 MW m$^{-2}$ K$^{-1}$, while the TBC values of the SiO$_x$-SiO$_x$ interface can be regarded as greater than 150 MW m$^{-2}$ K$^{-1}$. The excessively high fitted values are artifacts arising from the extremely high SiO$_x$-SiO$_x$ TBC coupled with very low sensitivity, and thus these values essentially lack meaning of physics.

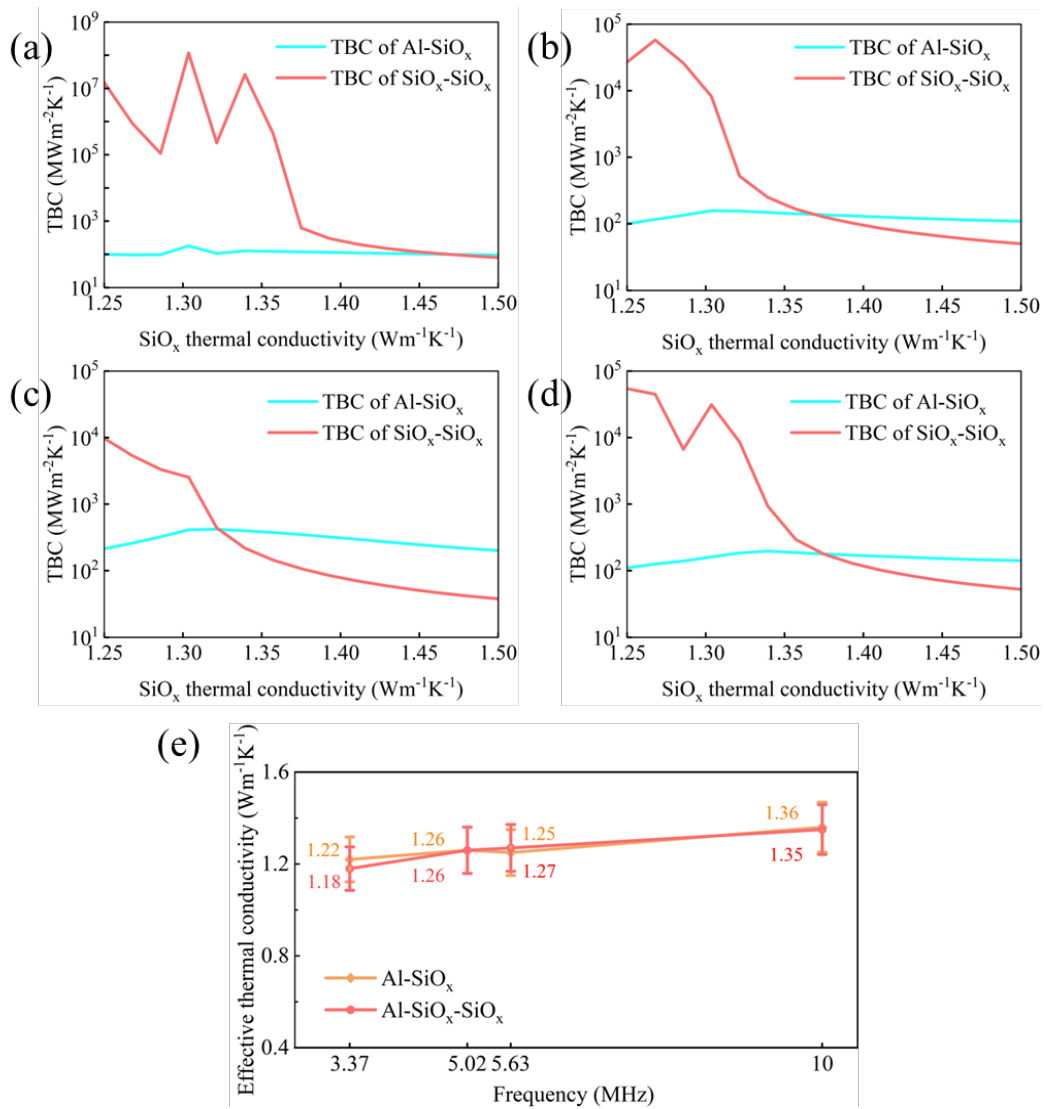

Figure 6. (a)–(d) Simultaneous fitting of Al-SiO$_x$ and SiO$_x$-SiO$_x$ interfacial TBC with fixed SiO$_x$ thermal conductivity. (e) Effective thermal conductivity results under assumed two-layer structure. The error bars represent an 8% uncertainty. Al-SiO$_x$ TBC is fixed at 117 MW m$^{-2}$ K$^{-1}$.

We fitted the effective thermal conductivity of the structure with the introduced SiO$_x$-SiO$_x$ interface by simplifying the Al-SiO$_x$-SiO$_x$ trilayer structure into an Al-SiO$_x$ bilayer structure. As shown in Fig. 6(e), the effective thermal conductivity derived from this bilayer simplification exhibits almost no difference from the fitting results obtained from regions without the SiO$_x$-SiO$_x$ interface, which is consistent with the nearly indistinguishable ratio curves. All the equivalent fitting results are distributed within an 8% error range relative to 1.25 W m$^{-1}$ K$^{-1}$, and the slight discrepancies in the results at different modulation frequencies can be attributed to measurement errors. These results demonstrate that the SiO$_x$-SiO$_x$ dielectric interface forms a high-quality interfacial structure after bonding and low-temperature annealing. The formation of Si–O–Si covalent bonds improves the interfacial quality, leading to an extremely short Kapitza length corresponding to the interfacial TBC. Thus, the interfacial thermal resistance introduced by the SiO$_x$-SiO$_x$ interface is negligible. This work provides insights into the fabrication process and thermal properties of dielectric-dielectric interfaces for hybrid bonding applications.

## AUTHOR DECLARATIONS

### Conflict of Interest

The authors have no conflicts to disclose.

**Author Contributions**

**Xingqiang Zhang:** Data curation (equal); Formal analysis (equal); Investigation (equal); Methodology (equal); Validation (equal); Visualization (equal); Writing – original draft (lead); Writing review & editing (equal). **Liu Chang:** Data curation (equal); Formal analysis (equal); Investigation (equal); Writing review & editing (equal). **Liyi Li:** Conceptualization (equal); Methodology (equal); Resources (equal); Funding acquisition (equal); Supervision (equal). **Zhe Cheng:** Conceptualization (equal); Data curation (equal); Funding acquisition (equal); Investigation (equal); Methodology (equal); Project administration (equal); Resources (equal); Supervision (equal); Visualization (equal); Writing– review & editing (equal).

**DATA AVAILABILITY**

The data that support the findings of this study are available from the corresponding author upon reasonable request.

**ACKNOWLEDGEMENTS**

The authors acknowledge the financial support from the National Key Research and Development Program of China (Grant No. 2024YFA1207900) and the National Natural Science Foundation of China (NSFC) (Grant Nos. 62574007, T2550270, 92373204).